# Nucleosynthesis in Hot and Dense Media

**Samina S. Masood**
Department of Physics, University of Houston Clear Lake, Houston, USA
Email: Masood@uhcl.edu

**Abstract**

We study the finite temperature and density effects on beta decay rates to compute their contributions to nucleosynthesis. QED type corrections to beta decay from the hot and dense background are estimated in terms of the statistical corrections to the self-mass of an electron. For this purpose, we re-examine the hot and dense background contributions to the electron mass and compute its effect to the beta decay rate, helium yield, energy density of the universe as well as the change in neutrino temperature from the first order contribution to the self-mass of electrons during these processes. We explicitly show that the thermal contribution to the helium abundance at T = m of a cooling universe 0.045 % is higher than the corresponding contribution to helium abundance of a heating universe 0.031% due to the existence of hot fermions before the beginning of nucleosynthesis and their absence after the nucleosynthesis, in the early universe. Thermal contribution to helium abundance was a simple quadratic function of temperature, before and after the nucleosynthesis. However, this quadratic behavior was not the same before the decoupling temperature due to weak interactions; so the nucleosynthesis did not even start before the universe had cooled down to the neutrino decoupling temperatures and QED became a dominant theory. It is also explicitly shown that the chemical potential in the core of supermassive and superdense stars affect beta decay and their helium abundance but the background contributions depend on the ratio between temperature and chemical potential and not the chemical potential or temperature only. It has been noticed that temperature plays a role of regulating parameter in an extremely dense systems.

## 1. Introduction

Primordial nucleosynthesis was facilitated by means of beta decay processes. When nuclear formation takes place inside a hot and dense medium, nucleosynthesis parameters are affected by the finite temperature and density (FTD) of the background due to the modifications in the beta decay rates in a statistical medium. Beta decay rates are pronounced in a different manner; they also contributed differently to primordial nucleosynthesis in the early universe and inside hot and dense stellar cores because of the difference in the corresponding medium properties. Major contributions came from the physical mass of electron and the phase space for beta processes.

Standard Big Bang Model (SBBM) of the universe [1] indicates that the universe went through nucleosynthesis when it cooled to $10^{10}$ K. Beta decay processes started when the baryon density $\eta_B$ was as low as $10^{-10}$; Big Bang Nucleosynthesis (BBN) in the universe started around the same time. It is known that the beta decay rates [1-3] depend on masses of particles and the phase space. Field theory of quantum electrodynamics (QED) at Finite Temperature and Density (FTD) shows that the electron mass, wave function and charge are modified in a hot and dense medium. Changes in the physical properties of electrons are determined through their interaction with particles in a hot and dense medium while they propagate through the medium. These physically measureable values of parameters correspond to the effective parameters of the theory in that medium.

SBBM of the universe, the most well-known model of cosmology, predicts the abundances of light elements in the early universe. Beta decay processes are observed on Earth as well as in the extremely hot universe and inside the extremely hot and dense stellar cores, whereas the nucleosynthesis only occurs under the special conditions of temperatures and densities such as in the primordial universe and in the stellar cores. Since beta decay is a weak process and nucleosynthesis is expected to start after the decoupling temperature [4,5], QED is not enough to fully describe nucleosynthesis. One can think about a major role of electroweak theory at FTD in nucleosynthesis. However, the concentrations of hot and dense W's and Z's in the background are totally insignificant when compared to photons. The electroweak corrections of the background are suppressed by the heavy masses of electroweak mediators W and Z.

Neutrino mass being tiny enough (even if it exists in the extended standard models [4-6]) is ignored for all practical purposes. However, some of the corrections due to the form factors of neutrinos [7-11] cannot be ignored at these temperatures and densities. Although, for extremely high temperatures at the beginning of the universe, we may not be able to totally ignore the electroweak background corrections [12] for the study of leptogenesis, but at that point, extensions of standard models have to be used (See for example: [12-15]).

Beta decay processes are studied in a hot and dense media to accurately calculate their contributions to nucleosynthesis. However, all nucleosynthesis temperatures are well below the electroweak scale and we can easily ignore thermal contributions to electroweak processes in this range. If the neutrino is not massless and we use minimal extension of the standard model to work with the tiny mass of Dirac type neutrino, the weak processes have to be incorporated. Also the properties of neutrinos are significantly modified in hot and dense media at this scale, provided the neutrinos have nonzero mass due to extensions in the standard model, we would have to include thermal contributions due to the nonzero mass of neutrino in the minimal extensions (or other extensions) of the standard model. However, in this paper we just restrict ourselves to the standard electroweak model with the massless neutrino and exclusively study the QED type FTD corrections [16-30] only. For this purpose, we consider previously calculated relationships of electron mass with nucleosynthesis parameters, such as beta decay rate and helium abundance in the early universe [1-3].

High energy physics provides a theoretical justification of the SBBM. When the universe was less than a second old, it was extremely hot and electron-positron pairs were created as the first matter particles. Properties of electrons in the very early environment of the universe were not the same as they are in a vacuum. The behavior of electrons in the very early universe can be predicted incorporating thermal background effects on the physical properties of electrons. We use the renormalization scheme of QED to determine the renormalization constants of QED, such as electron self-mass, charge and wave function, to study the physical properties of electrons at high temperatures.

Renormalization of QED at finite temperature and density ensures a divergence free QED in hot and dense media. The most general calculations of the first order thermal loop corrections to electron self-mass, charge and wave function are performed in detail, incorporating the background density effects through the chemical potential [27-32]. Calculations of the second order corrections, at finite temperatures [25,26], to the renormalization constants, in different ranges of temperatures are already there in literature. However, it is not possible to separate thermal corrections from the vacuum contributions. Also the validity of the renormalization scheme fully justifies that the second order corrections are significantly small as compared to the first order thermal corrections.

In the next section we briefly mention the calculational scheme and rewrite some of the self-mass of electron expressions in more useful form for the more relevant regions of temperature and chemical potentials of astrophysical systems.

Section 3 is devoted to a discussion of nucleosynthesis at finite temperature and density corrections to the electron mass in the early universe and for the stellar cores. Just for simplicity, we have not included the effect of strong magnetic fields in the core of neutron stars as it has to be studied separately, in detail, because of the complexity of the issue. Also, interestingly, due to the recently observed existence [33] of superfluidity, a comprehensive study of this aspect of the problem is demanded. High abundance of helium at high density and low temperature, in the presence of strong magnetic fields may provide favorable conditions for superfluidity.

## 2. Calculational Scheme

We summarize the previously calculated results of the renormalization constants of QED in the real-time formalism, up to the one loop level, at FTD. It is possible to separate out the temperature dependent contributions from the vacuum contributions as the statistical distribution functions contribute additional statistical terms both to fermion and boson propagators in the form of the Fermi-Dirac distribution and the Bose-Einstein distribution functions, respectively. The Feynman rules of vacuum theory are used with the statistically corrected propagators [16-21] given as

$$D(k) = \frac{i}{k^2} + 2\pi n_B(k_0)\delta(k^2), \quad (1)$$

$$n_B(k_0) = \frac{1}{k^2},$$

and the fermion propagator is defined as

$$S_F(p) = \frac{i}{\slashed{p} - m + i\varepsilon} - 2\pi\delta(p^2 - m^2)\left[\theta(p_0)n_F^+(p,\mu) + \theta(-p_0)n_F^-(p,\mu)\right] \quad (2)$$

where the corresponding energies of electrons (positrons) are defined as

$$E_{p,k} = \sqrt{(\vec{p},\vec{k})^2 - m^2}.$$

$\mu$ is the chemical potential which is assigned a positive sign for particles and negative sign for antiparticles. Fermion distribution function at FTD can then be written as

$$n_F^\pm \equiv n_F(p \pm \mu) = \frac{1}{e^{-\beta(|E_p|\pm\mu)} + 1},$$

where the positive sign corresponds to electron and negative to positron. This sign difference (in $n_F^\pm(p,\mu)$) determines the difference in behavior of particles (antiparticles) in a dense background. It is obvious from Eqs. (1) and (2) that the photons and electrons propagated differently, in the beginning of the hot early universe, right after its creation. Photons, being massless particles, exhibit zero chemical potential and no density effects, whereas the electrons (positrons) propagation in the medium help to understand several issues in dense media which are out of scope of this paper.

The electron mass, wave function and charge are then calculated in a statistical medium using Feynman rules of QED, using the modified propagators given in Eqs. (1) and (2). The renormalization constants are evaluated for different hot and dense systems to understand the propagation of electrons in such media. These renormalization constants behave as effective parameters of hot and dense systems. We briefly overview the calculations of the relevant parameters of QED at FTD and explain how the renormalization constants can be used to describe the physical behavior of the hot and dense systems.

## 2.1. Self-mass of Electron

The renormalized mass of electrons $m_R$ can be represented as a physical mass $m_{phys}$ of electron and is defined in a hot and dense medium as,

$$m_R \equiv m_{phys} = m + \delta m(T=0) + \delta m(T,\mu). \quad (3)$$

where $m$ is the electrons mass at zero temperature; $\delta m(T=0)$ represents the radiative corrections from vacuum and $\delta m(T,\mu)$ is the perturbative corrections to the mass (self-mass) due to its interaction to the statistical background at FTD. Thermal effects are computed by means of the particle interaction with the hot particles of the medium at temperature T; the density effects are estimated in terms of the chemical potential $\mu$. The physical mass can get radiative corrections at different orders of $\alpha$ and can be written as

$$m_{phys} \cong m + \delta m^{(1)} + \delta m^{(2)}, \quad (4)$$

where $\delta m(1)$ and $\delta m(2)$ are the shifts in the electron mass in the first and second order in $\alpha$, respectively. This perturbative series can go to any order of $\alpha$, as long as it is convergent. The physical mass is calculated by locating the pole of the fermion propagator:

$$\frac{i(\not{p}+m)}{p^2 - m^2 + i\varepsilon}$$

in thermal background. For this purpose, we sum over all the same order diagrams at FTD. Renormalization is established by demonstrating the order-by-order cancellation of singularities at finite temperatures and densities. All the terms from the same order in $\alpha$ are combined together to evaluate the same order contribution to the physical mass given in Eq.(4) and are required to be finite to ensure order by order cancellation of singularities. The physical mass in thermal background up to order $\alpha^2$ [23-27] is calculated at finite temperature, using the renormalization techniques of QED. Higher order background contributions to electron mass, due to the chemical potential are still to be computed. However, following the renormalization scheme of vacuum, we may write the self-mass term as,

$$\sum(p) = A(p) E\gamma_0 - B(p)\vec{p}.\vec{\gamma} - C(p), \quad (5)$$

where A(p), B(p), and C(p) are the relevant coefficients and are modified at FTD. Taking the inverse of the propagator where the momentum and mass terms separated as,

$$S^{-1}(p) = (1-A) E\gamma^0 - (1-B) p.\gamma - (m-C). \quad (6)$$

The temperature-dependent radiative corrections to the electron mass are obtained from the FTD propagators in Eq.(6). These corrections are rewritten in terms of the boson loop integral I's and the fermion loop integrals J's with the one loop level as,

$$E^2 - |\vec{p}|^2 = m^2 + \frac{\alpha}{2\pi^2}(I.p + J_B.p + m^2 J_A) \equiv m_{phys}^2, \quad (7)$$

with

$$I_A = 8\pi \int \frac{dk}{k} n_B(k),$$

$$I^\mu = 2\int \frac{d^3k}{k} n_B(k)\frac{k^\mu}{p_\nu k^\nu} = 2\int \frac{d^3k}{k} n_B(k)\frac{(k_0,\vec{k})}{E_p k_0 - \vec{p}.\vec{k}},$$

$$J_A = \int \frac{d^3l}{E_l} n_F(E_l,\pm\mu)\left[\frac{1}{E_p E_l + m^2 - \vec{p}.\vec{l}} - \frac{1}{E_p E_l - m^2 - \vec{p}.\vec{l}}\right],$$

$$J_B^\mu = \int \frac{d^3l}{E_l} n_F(E_l,\pm\mu)\left[\frac{(E_p+E_l,\vec{p}+\vec{l})}{E_p E_l + m^2 - \vec{p}.\vec{l}} - \frac{(E_p-E_l,\vec{p}+\vec{l})}{E_p E_l - m^2 - \vec{p}.\vec{l}}\right]. \quad (8)$$

$$I.p = \frac{4\pi^3 T^2}{3}, \quad (9)$$

and

$$J_B \cdot p = 8\pi \left[ I_1(m\beta, \pm\mu) - \frac{m^2}{2} I_2(m\beta, \pm\mu) \right] \quad (10)$$

$$= 8\pi \left[ \frac{m}{\beta} a(m\beta, \pm\mu) - \frac{m^2}{2} b(m\beta, \pm\mu) - \frac{1}{\beta^2} c(m\beta, \pm\mu) \right].$$

$$m_{phys}^2 = m^2 \left[ 1 - \frac{6\alpha}{\pi} b(m\beta, \pm\mu) \right]$$
$$+ \frac{4\alpha}{\pi} \left[ mT a(m\beta, \pm\mu) + \frac{2}{3} \alpha\pi T^2 - \frac{6}{\pi^2} c(m\beta, \pm\mu) \right]. \quad (11)$$

$$\frac{\delta m}{m} \simeq \frac{1}{2m^2}(m_{phys}^2 - m^2) \simeq \frac{\alpha\pi T^2}{3m^2} \left[ 1 - \frac{6}{\pi^2} c(m\beta, \pm\mu) \right]$$
$$+ \frac{2\alpha}{\pi} \frac{T}{m} a(m\beta, \pm\mu) - \frac{3\alpha}{\pi} b(m\beta, \pm\mu). \quad (12)$$

where $+\mu(-\mu)$ correspond to the chemical potential of electron (positron) and correspond to the density of the system. $\delta m/m$ is the relative shift in electron (positron) mass due to finite temperature and density of the medium, determined in Ref. [19] with

$$a(m\beta, \pm\mu) = \ln(1 + e^{-\beta(m\pm\mu)}),$$
$$b(m\beta, \pm\mu) = \sum_{n=1}^{\infty} (-1)^n e^{\mp\beta\mu} Ei(-nm\beta),$$
$$c(m\beta, \pm\mu) = \sum_{n=1}^{\infty} (-1)^n \frac{e^{-n\beta(m\pm\mu)}}{n^2}. \quad (13)$$

The validity of Eq.(13) can be ensured for $T \leq 2$ MeV in the early universe where $|\mu|$ is ignorable (see Ref.[8], for details). The big bang theory of cosmology and all the observational data of the universe agree that primordial nucleosynthesis occurred when the universe cooled down to around $10^{10}$K. On the other hand, renormalization of quantum electrodynamics (QED) in hot and dense media indicate that the hot medium contribution at T= m is different for a heating and a cooling system [4,5]. This disagreement supports the big bang model as it indicates that the universe was going through nucleosynthesis at those temperatures and the compositional change in the universe.

The convergence of Eq.(4) can be ensured [4,5] at T $\leq$ 2MeV as $\delta m/m$ is sufficiently smaller than unity within this limit[17]. This scheme of calculations will not work for higher temperatures and the first order corrections may exceed the original values of QED parameters, after 2MeV. At low temperature T < m, the functions a(m$\beta$,±$\mu$), b(m$\beta$,±$\mu$), and c(m$\beta$,±$\mu$) fall off in powers of $e^{-m\beta}$ in comparison with $(T/m)^2$ when ($\mu$<m<T) and can be neglected in the low temperature limit giving,

$$\frac{\delta m}{m} \to \frac{\alpha\pi T^2}{3m^2}. \quad for\ T \ll m \quad (14)$$

In the high-temperature limit, neglecting $\mu$, a(m$\beta$,±$\mu$) and b(m$\beta$,±$\mu$) are still vanishingly small whereas

c(m$\beta$,±$\mu$) $\to$ $-\pi^2/12$, yields

$$\frac{\delta m}{m} \to \frac{\alpha\pi T^2}{2m^2}. \quad for\ T \gg m \quad (15)$$

The above equations give $\delta m/m$ = 7.647×10−3 $T^2/m^2$ for low temperature and $\delta m/m$ = 1.147×10−2 $T^2/m^2$ for high temperature, showing that the rate of change of mass $\delta m/m$ is larger at T > m as compared to T < m. Subtracting eq. (12) from (13), the change in $\delta m/m$ between low and high temperature ranges can be written as

$$\Delta\left(\frac{\delta m}{m}\right) = \pm\frac{\alpha\pi T^2}{6m^2} = \pm 3.8 \times 10^{-3} \frac{T^2}{m^2}. \quad (16)$$

Nucleosynthesis is held responsible for that. Eqs. (14) and (15) show that thermal corrections to the electron self-mass are expressed in terms of T/m both for low T and for high T. It is only during the nucleosynthesis that self-mass deviates from the T/m and has to be expressed in terms of $a_i$ functions derived by Masood [31, 32]. Actually since there is no significant change in the density of electrons or photons, the thermal contributions to the beta decay and other nucleosynthesis parameters, including the helium yield, do not change much as long as the low T or high T approximations are valid to use Eqs. (14,15). However at the higher loop level, the low T and high T approximations are not so well described and they are tied up with vacuum effects. Moreover, the self-mass expressions are much more complicated to retrieve thermal corrections to helium yield and the above given relations between the self-mass and the nucleosynthesis parameters have to be revised as the phase space for beta processes changes at higher loops and will involve Boltzmann Equations. So we restrict ourselves to the one loop corrections as they are the only relevant corrections at such temperatures. However, Eqs. (14) and (15) do not coincide at T=m, though they are both derived from the same master equation, i.e; Eq.(12). This discrepancy has a physical justification that the nucleosynthesis starts around T~m, and the parameters of the theory become complicated function of temperature during nucleosynthesis. It therefore matters a lot if the system has gone through nucleosynthesis (a cooling system) or it is starting nucleosynthesis (the heating system).

When the density effects are not ignorable, the perturbative series is still valid at much higher temperatures, due to the reason that the growth of mass is slowed down significantly in a dense system. High densities and smaller mean free paths automatically ensure the validity of the perturbative expansion as the argument of the exponential changes from $\beta m$ to $\beta(m\pm\mu)$ and the expansion parameter changes from $m/T$ to $(m\pm\mu)/T$ as $T = 1/\beta$, for such systems. When the chemical potential is large, it can overcome thermal effects as we deal with $\beta\mu$ and not $\beta m$, the expansion parameters. Especially for electrons (at $\mu \gg m \geq T$), the distribution function $n^{\pm}_F(p,\mu)$ reduces to the function $\theta(\mu - E_l)$ for electrons, (and vanishes for positrons) providing $\mu$ as an upper limit to $E_l$, such that the integration is simplified as,

$$\int_m^\infty \frac{dE_l}{E_l}\theta(\mu - E_l) = \ln\frac{\mu}{m},$$

$$\int_m^\infty E_l dE_l \theta(\mu - E_l) = \frac{1}{2}(\mu^2 - m^2),$$

$$\int_m^\infty \frac{dE_l}{E_l^3}\theta(\mu - E_l) = -\frac{1}{2}\left(\frac{1}{\mu^2} - \frac{1}{m^2}\right). \quad (17)$$

Inserting the results of integrations of Eq. (17) into Eqs. (8) and (9), we obtain

$$J_A \simeq -8\pi \ln\frac{\mu}{m} + 8\pi \frac{E_p^2}{\mu^2}\left(1 - \frac{\mu^2}{m^2}\right),$$

$$\int_0^\infty \frac{l^2 dl}{E_l}\theta(\mu - E_l) = -\frac{m^2}{2}\ln\frac{\mu}{m} - \frac{m^2}{2}\left(1 - \frac{\mu^2}{m^2}\right)\left(1 - \frac{m^2}{\mu^2}\right)$$

$$\frac{J_B^0}{E} \simeq \pi\left(1 - \frac{\mu^2}{m^2}\right)\left(\frac{2m^2}{pE}\ln\frac{1-v}{1+v} - \frac{E^2}{4m^2}\right) - 4\pi\ln\frac{\mu}{m}, \quad (18)$$

Giving

$$m_{phys}^2 \simeq m^2 - \frac{6\alpha}{\pi}\ln\frac{\mu}{m} + \frac{2\alpha}{\pi}m^2\left(1 - \frac{\mu^2}{m^2} + \frac{2p^2}{\mu^2} - 1\right), \quad (19)$$

and

$$\frac{\delta m}{m}(T,\mu) \simeq -\frac{3\alpha}{\pi}\ln\frac{\mu}{m}$$

$$+ \frac{\alpha}{\pi}\left(1 - \frac{\mu^2}{m^2}\right)\left(3\frac{m^2}{\mu^2} + \frac{2p^2}{\mu^2} - 1\right). \quad (20)$$

Eqs. (19) and (20) give the electron self-mass for the extremely dense stellar cores which have very high temperatures, but due to the extremely dense situation, cannot be treated as purely hot systems. Neutron stars provide a good example of such systems. However, in neutron stars, high magnetic field effects are not ignorable either, though they are out of the scope of this paper. Eq. (20) shows that the extremely large values of $\mu$ will lead to the dominant behavior of electron mass as

$$\frac{\delta m}{m}(T,\mu) \simeq \frac{\alpha}{\pi}\frac{\mu^2}{m^2}. \quad (21)$$

Eq. (21) shows the mass dependence on chemical potential is just as it were at T, in the extremely large chemical potential values and self-mass of electron grows as $\mu^2/m^2$ and can be plotted as in Figure 1.

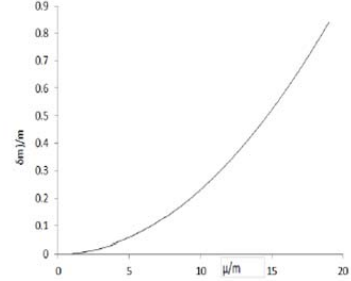

**Figure 1.** *Electron self-mass as a function of chemical potential. for $\mu > T > m$*

Using the electron self-mass contribution from FTD background, we can calculate the background contribution to the helium abundance parameter, corresponding to the astrophysical systems of interest. This contribution is small, still non-ignorable.

## 3. Beta Decay and Nucleosynthesis

The big bang model of the early universe indicates that nucleosynthesis takes place around the temperatures of the electron mass, that is around $10^{10}$ K. Nucleosynthesis was initiated by the creation of protons as hydrogen nuclei. A proton can capture an electron to create neutrons which can decay back to a proton through beta decay. Therefore, the beta decay processes are usually studied in detail to understand the start of nucleosynthesis. The abundances of light elements are related to the neutron to proton ratio as well as nucleon to photon ratio, calculated in the medium at the time of nucleosynthesis. Beta-processes kept the ratio between protons and neutrons in all the relevant channels and photons **were** regulated by the background temperatures [1-3].

$$\Delta Y = 0.2 \frac{\Delta \tau}{\tau} = -0.2 \frac{\Delta \lambda}{\lambda}, \quad (22)$$

where $\Delta\tau/\tau$ is relative change in neutron half-life and $\Delta\lambda/\lambda$ is relative change in neutron decay rate.

### 3.1. First Order Contributions from Electron Selfmass

Neutron decay rate or half-life can easily be related to the electron mass. It has been explicitly shown [1-3] that the radiative corrections, especially thermal background contributions to beta decay rate and all the other parameters can be expressed in terms of the self-mass of electron. So the radiative corrections to beta decay rate can be expressed as

$$\frac{\Delta \lambda}{\lambda} = -0.2 \left(\frac{m}{T}\right)^2 \frac{\delta m}{m}, \quad (23)$$

with m as the mass of the propagating electron, T is the temperature of background heat bath and $\delta m/m$ is the radiative corrections to electron mass due to its interaction with thermal medium. In the early universe the temperature effects were dominant with ignorable density effects as the chemical potential μ of the particles satisfies the condition $\mu/T \leq 10^{-9}$. The neutrino temperature $T_\nu$ can be written as

$$\frac{\Delta T_\nu}{T_\nu} = -0.1 \left(\frac{m}{T}\right)^2 \frac{\delta m}{m},$$

$$\frac{\Delta T_\nu}{T} = -\left(\frac{m}{5T}\right)^2 \frac{\delta m}{m}.$$

Considering all of the three generation of leptons, we can express ΔY,

$$\Delta Y \simeq -0.01 \frac{\Delta T_{\nu_e}}{T_{\nu_e}} + 0.04 \frac{\Delta T_{\nu_\mu}}{T_{\nu_\mu}} + 0.04 \frac{\Delta T_{\nu_\tau}}{T_{\nu_\tau}}. \quad (24)$$

working below those temperatures in the early universe.

$$\frac{\Delta \rho_T}{\rho_T} = -\left(\frac{m}{4T}\right)^2 \frac{\delta m}{m}. \quad (25)$$

The total energy density $\rho_T$ of the universe affects the expansion rate of the universe H

$$H = \left(\frac{8}{3}\pi G \rho_T\right)^{\frac{1}{2}} \quad (26)$$

giving

$$H = \left(\frac{8}{3}\pi G \rho_T \left(1 - \left(\frac{m}{4T}\right)^2 \frac{\delta m}{m}\right)\right)^{\frac{1}{2}} \quad (27)$$

which corresponds to the change in H as

$$\frac{\Delta H}{H} \simeq -0.5 \left(\frac{m}{4T}\right)^2 \frac{\delta m}{m} \quad (28)$$

and Eq. (22) can be re-written as

$$\Delta Y = 0.4 \left(\frac{m}{T}\right)^2 \frac{\delta m}{m}. \quad (29)$$

Thermal contribution to beta decay for T ≤ m is -0.00153 and at T ≥ m is -0.00229, whereas the contribution to helium synthesis parameter is 0.000306 and for large T it is 0.000459. It gives thermal corrections to Y for a heating universe: 0.03% and for a cooling universe it is 0.045 % of the accepted value of around 0.25. Figure 2 gives a comparison of helium abundance parameter Y as a function of temperature before and after nucleosynthesis.

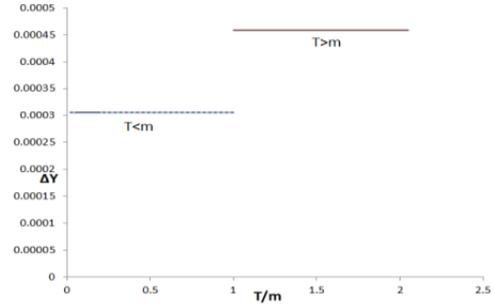

**Figure 2.** *A comparison of helium abundance at low temperature and the helium abundance at high temperature.*

T ~ m range of temperature is particularly interesting from the point of view of primordial nucleosynthesis. It has been found that some parameters in the early universe such as the energy density and the helium abundance parameter Y become a slowly varying function of temperature [2]; whereas they remain constant before and after the nucleosynthesis as the quadratic term in self-mass contribution from the background cancels out when Eqs. (14) and (15) are substituted. The chemical potential self-mass contributions can also be calculated from Eq. (21) directly as

$$\Delta Y = 0.04 \left(\frac{m}{T}\right)^2 \frac{\alpha}{\pi} \frac{\mu^2}{m^2}. \tag{30}$$

for the chemical potential μ sufficiently greater than temperature T as well as the electron mass. T < m and μ > m. However, it can be seen that the presence of (m/T) factor in Eq. (30) ensures that the helium synthesis will blow up at low temperature or the system will not maintain equilibrium at low T. Therefore, temperature acts like a regulating parameter at high chemical potential indicating that inside the stellar cores with large chemical potential of electron, the temperature has to be high. Figure 3 gives a plot of helium abundance as a function of chemical potential at constant temperature. When T reduces from 0.25 MeV to 0.05 MeV, the helium abundance contribution (due to beta decay) changes tremendously for the same chemical potentials. Therefore the large chemical potential and high density effects are the same at high temperatures as long as the chemical potential is sufficiently higher than the temperature. In a way, the fluidity of helium depends on a ratio (μ / T) and not on T or μ only and the magnetic field. However, we ignore the effects of magnetic field in this paper.

## 3.2. Second Order Contributions from Electron Selfmass

Study of the early universe has passed through the stage of theoretical prediction to observations and testing. COBE (Cosmic Background Explorer), WMAP (Wilkinsin Microwave Anisotropy Probe) and LHC (Large Hadron Collider) provide data [34-39] to test the standard model of cosmology and the particle interaction theories to much better precision level than earlier. So the precise calculations are needed to correlate observational data with theoretical models and use data to test models; or get help from theoretical models to develop techniques to make observations more precise. So the first order results may not be accurate enough to meet the precision level of future probes and we need to go to second order of perturbation. However, all of the previously established relations between nucleosynthesis parameters and the electron mass are acceptable approximations for the first order corrections used in the last section. Also FTD corrections can be studied independently at the one loop level only where it is possible to segregate between radiative corrections from vacuum and FTD corrections due to the interactions with the statistical background, in real time formalism. The validity of

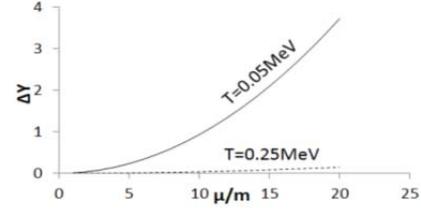

**Figure 3** *Plot of helium yield versus a function of chemical potential, when temperature is smaller than the chemical potential. Dotted line is a plot for T/m=0.1 and the broken line gives same plat for T/m=0.5 showing that the larger temperature will give smaller contributions to helium abundance at high densities.*

these expressions for the higher loops is questionable as the overlap between the hot and cold loops increase the contributions of hot terms and they are not totally different from cold contributions. So we cannot calculate the second order thermal

$$\Delta Y = 0.04 \left(\frac{m}{T}\right)^2 \left(\frac{1}{3}\frac{\alpha\pi T^2}{m^2} + 15\alpha^2\left(\frac{T^2}{m^2}\right)\right), \tag{31}$$

out at the two loop level. When we go to check the effects of higher order background contributions, we may need to re-write or at least re-check the validity of these relations. Moreover, thermal contributions to electron mass during nucleosynthesis are so complicated [22-23] that even to extract the correct thermal behavior, simple analytical methods cannot be used and numerical evaluation of hot terms will be needed. Just for comparison between one-loop and two loop behavior; we mention existing approximate results [26]. Using the second order contributions to the electron mass at low temperature (T < m), leading order contributions to the helium yield can be computed as [37]

Whereas, the leading order contributions at high temperature (T > m) comes out to be

$$\Delta Y = 0.04 \left(\frac{m}{T}\right)^2 \left(\frac{1}{2}\frac{\alpha\pi T^2}{m^2}\right. \tag{32}$$
$$\left. + \frac{\alpha^2\pi^2}{4}\left(\frac{T^2}{m^2}\right)^2 - \frac{\alpha^2}{4}\frac{m^2}{T^2}\right).$$

Eqs. (31) and (32) correspond to second order corrections, in these approximate methods. Eq. (31) shows that the second order corrections are sufficiently smaller than the first order corrections at low temperature given as $3.38 \times 10^{-4}$ in comparison with the one loop low temperature contribution of $3.06 \times 10^{-4}$. However, the high temperature contributions from Eq. (32) are not very encouraging as they not only reduce the helium abundance at high temperature, the last term on the right hand side of equation induces very strong temperature dependence at lower temperatures and helium yield becomes negative before the nucleosynthesis is started. This unusual behavior has to be carefully studied, even before we look for its physical interpretation.

## 4. Results and Discussions

Nucleosynthesis plays an important role in understanding the astrophysical problems such as the matter creation in the early universe, inflation, energy density of the universe and stellar structure formation. Primordial nucleosynthesis in a very hot and extremely low density universe was significant until the production of $^4$He and has been studied in detail, not only to resolve some key issues of SBBM of cosmology; but also some important issues of nuclear interactions in high energy physics. Study of nucleosynthesis also helps to understand large scale structure formation, energy density of the universe and chemical evolution of galaxies. There are theoretical as well as observational predictions for primordial $^4$He yield. Beta decay and weak interactions played an important role during the primordial nucleosynthesis, until it froze. Afterward the temperature was lowered and the available neutrons fused to form the light nuclei. The $^4$He abundance parameter is however sensitive to the electron mass and the temperature dependence of phase space due to the Fermi-Dirac distribution function for hot electrons and Bose-Einstein distribution function for the bosons. However, the radiative corrections are not very important as we do not have to include radiative corrections to the decay rates [3,32], at least at the one-loop level. Therefore, the existence of finite temperature QED background makes it relevant to include its corrections to electron propagator. Background corrections mainly appear from the self-mass of electron. In the previous section, we calculated FTD corrections to different parameters of cosmology in terms of δm/m. We list the numerical values of these parameters as a function of temperature. **Tables 1 and 2** indicate the numerical values of all these parameters and show the difference of values of these parameters at T=m for a cooling and a heating system. It is also noticed that regardless of change in the electron selfmass, the parameters during nucleosynthesis does not change much with temperature as nucleosynthesis parameters are slowly varying function of temperature however fall in temperature was very fast in the early universe.

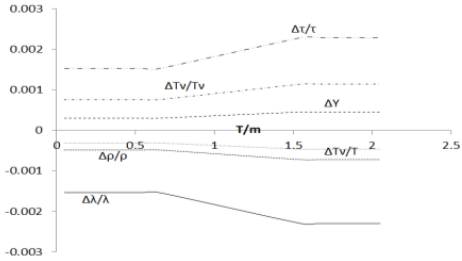

**Figure 4.** *Plot of different Nucleosynthesis parameters as a function of temperature.*

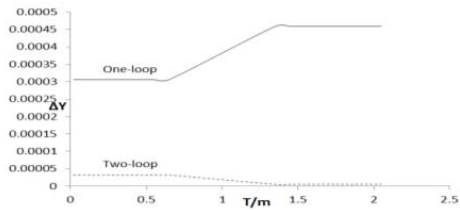

**Figure 5.** *Comparison of one-loop and two loop level contributions.*

**Table 1**. *Numerical values ($\times 10^{-3}$) of parameters corresponding to $T \leq m$.*

| T/m | δm/m | Δλ/λ | ΔY | Δτ/τ | $\Delta T_\nu/T_\nu$ | $\Delta T_\nu T$ | $\Delta \rho_T/\rho_T$ |
|---|---|---|---|---|---|---|---|
| | | | Low Temperature values | | | | |
| 0.05 | 0.19 | -1.53 | 0.31 | 1.53 | 0.765 | -0.31 | -0.478 |
| 0.15 | 0.17 | -1.53 | 0.31 | 1.53 | 0.765 | -0.31 | -0.478 |
| 0.25 | 0.48 | -1.53 | 0.31 | 1.53 | 0.765 | -0.31 | -0.478 |
| 0.35 | 0.93 | -1.53 | 0.31 | 1.53 | 0.765 | -0.31 | -0.478 |
| 0.45 | 1.55 | -1.53 | 0.31 | 1.53 | 0.765 | -0.31 | -0.478 |
| 0.55 | 2.31 | -1.53 | 0.31 | 1.53 | 0.765 | -0.31 | -0.478 |
| 0.65 | 3.23 | -1.53 | 0.31 | 1.53 | 0.765 | -0.31 | -0.478 |
| 0.75 | 4.30 | -1.53 | 0.31 | 1.53 | 0.765 | -0.31 | -0.478 |
| 0.85 | 5.52 | -1.53 | 0.31 | 1.53 | 0.765 | -0.31 | -0.478 |
| 0.95 | 6.90 | -1.53 | 0.31 | 1.53 | 0.765 | -0.31 | -0.478 |
| 1.00 | 7.65 | -1.53 | 0.31 | 1.53 | 0.765 | -0.31 | -0.478 |

**Table 2**. *Numerical values ($\times 10^{-3}$) of parameters corresponding to $T \geq m$.*

| T/m | δm/m | Δλ/λ | ΔY | $\Delta T_\nu/T_\nu$ | Δτ/τ | $\Delta T_\nu/T$ | $\Delta \rho_T/\rho_T$ |
|---|---|---|---|---|---|---|---|
| | | | High Temperature Values | | | | |
| 1.00 | 11.47 | -2.3 | 0.46 | 1.147 | 2.3 | -0.46 | -0.717 |
| 1.05 | 12.64 | -2.3 | 0.46 | 1.147 | 2.3 | -0.46 | -0.717 |
| 1.15 | 15.16 | -2.3 | 0.46 | 1.147 | 2.3 | -0.46 | -0.717 |
| 1.25 | 17.92 | -2.3 | 0.46 | 1.147 | 2.3 | -0.46 | -0.717 |
| 1.35 | 20.90 | -2.3 | 0.46 | 1.147 | 2.3 | -0.46 | -0.717 |
| 1.45 | 24.12 | -2.3 | 0.46 | 1.147 | 2.3 | -0.46 | -0.717 |
| 1.55 | 27.56 | -2.3 | 0.46 | 1.147 | 2.3 | -0.46 | -0.717 |
| 1.65 | 31.23 | -2.3 | 0.46 | 1.147 | 2.3 | -0.46 | -0.717 |
| 1.75 | 35.12 | -2.3 | 0.46 | 1.147 | 2.3 | -0.46 | -0.717 |
| 1.85 | 39.26 | -2.3 | 0.46 | 1.147 | 2.3 | -0.46 | -0.717 |
| 1.95 | 43.61 | -2.3 | 0.46 | 1.147 | 2.3 | -0.46 | -0.717 |
| 2.05 | 48.20 | -2.3 | 0.46 | 1.147 | 2.3 | -0.46 | -0.717 |

We plot these parameters in Figure 4 as a function of temperature. It is clear from the above Figure that all of nucleosynthesis parameters become a slowly varying function of temperature during nucleosynthesis. They are constant before and after the nucleosynthesis. All of the nucleosynthesis parameters are plotted as a function of temperature at the one loop level. All of these parameters are constant for low and high temperatures and become slowly varying functions of temperatures during nucleosynthesis. The temperature and density dependence of these functions can be expressed in terms of a(mβ,±μ), b( mβ, ±μ) and c( mβ, ±μ) functions [3, 18, 19, 31, 32] in the corresponding ranges of temperature and density through δm/m given in Eqs. (14, 15 and 21), up to the one loop level in perturbation theory.

Figure 5 gives a comparison of two loop level selfmass corrections to nucleosynthesis and the one loop corrections. Since the two loop contributions are very contributed and even the asymptotic behavior gives a complicated expression so we just consider the disconnected graph to roughly estimate the second order selfmass correction to get an approximate comparison between the first order and second order contributions to helium abundance before and after the nucleosynthesis occurs. It is clear from Figure 5 that the higher order contributions to the nucleosynthesis are not so simple and cannot really be evaluated without numerical computations which we postpose for another paper. However, the second order behavior is expected to be similar to the first order perturbative correction. They are almost constant for low temperatures. However, during and after nucleosynthesis, their estimated approximate behavior given in Figure 5 may not be a good approximation. It can be easily seen from the graph that two-loop contribution is much smaller than one loop level and can easily be ignored. However, due to a negative term in Eq.(32) the helium synthesis seems to be decreasing rather than increasing at high temperatures. This leading order behavior has to be studied, in detail, as it may help to let us understand the universe in a little better way and or help to understand the limits on the validity of the theory or the calculational scheme. Numerical calculations of the thermal corrections at the higher loop level will be helpful at this stage.

The data from WMAP is still being interpreted and the later observational probes such as Planck [38] and James Webb Space Telescope [39] are expected to provide further fine tuning in precision values of these parameters. For this, even higher order modifications to electron mass at finite temperatures may provide further refinement to such corrections.